\documentclass[pra,twocolumn,showpacs,superscriptaddress,amsmath,amssymb]{revtex4}

\usepackage{graphicx}

\usepackage{color}

\begin{document}

\title{Dynamics of edge Majorana fermions in $\nu=\frac{5}2$ fractional quantum Hall effects}

\author{Yue Yu}
\affiliation{Institute of Theoretical Physics, Chinese
Academy of Sciences, P.O. Box 2735, Beijing 100080, China}
\date{\today}
\begin{abstract}
Commencing with the composite fermion description of the $\nu=5/2$
fractional quantum Hall effect, we study the dynamics of the edge
neutral Majorana fermions. We confirm that these neutral modes are
chiral and show that a conventional $p$-wave pairing interaction
between CFs does not contribute to the dynamics of the edge
neutral fermions. We find an important bilinear coupling between
the charged and neutral modes. We show that owing to this
coupling, the dispersion of the neutral modes is linear and their
velocities are proportional to the wave vector of the charged
mode. This dynamic origin of the motion of the edge Majorana
fermions was never expected before.
\end{abstract}


 \maketitle

\section {Introduction}

Edge states in quantum Hall effects(QHE) play an extremely
important role \cite{halp}. They are unique known media to reflect
the topological order of the bulk states in fractional QHE (FQHE)
\cite{wen}. Recently, the topological-protected quantum
computation \cite{kitaev} based on the possible non-abelian
statistics of quasiparticles in the filling factor $\nu=5/2$ FQH
(EFQH) states \cite{mr,GWW,note} has attracted great attentions
for it provides a possible candidate for the decoherence-free
qubit. Several experimental designs to detect the non-abelian
statistics of the $\nu=5/2$ EFQH state have been proposed
\cite{expd}. The crucial part of these designs was the
point-contact tunnelling of the quasiparticles between different
edges of the EFQH droplet.

The fundamental physics behind the non-abelian statistics in the
EFQH state is that low energy effective behaviors in the EFQH
system are controlled by a $k=2$ non-abelian Chern-Simons
topological quantum field theory in the bulk while the behaviors
of the edge states  by a $c=3k/(k+2)=3/2$ conformal field theory
which consists of chiral Majorana free fermionic modes with a
velocity $v_s$ and a chiral free bosonic mode with a velocity
$v_n\ll v_s$. Although this was already recognized by Moore and
Read \cite{mr} according to the correspondence between the wave
functions of various FQH states and the correlation functions of
the conformal field theory, a fundamental understanding from a
microscopic point of view still lacks. The velocity $v_n$ was not
determined. The chirality of the edge modes is a mystery because
the composite fermions (CFs) at the half-filling filling of the
Landau level do not see the effective magnetic field.

The present author and his co-workers have provided an effective
microscopic theory for the edge states of the odd denominator FQHE
\cite{yu}. We used a mean field theory to the CF Hamiltonian in
the bulk state. Project to a given Landau level, the electron band
mass is renormalized to the CF effective mass determined by the
Coulomb interaction. Shankar and Murthy have given an
understanding for this matter from a Hamiltonian formalism
\cite{sm}. A simplest calculation for the cancellation of the band
mass has been given in our previous work from a Hartree-Fock
approximation \cite{ysd}. However, the random phase approximation
calculation showed that the CF effective mass in the half-filled
Landau level is logarithmically divergent no matter what gauge is
taken \cite{HRL,ysd}.

On the other hand, if there is a pairing interaction, this
logarithmic divergence of the effective mass of vortex comes from
the extended states with their energy larger than the paring gap
\cite{ao}. The origin of divergence of the CF effective mass is
similar. Therefore, the physics with the energy scale lower than
the pairing gap can be studied in the same mean field
approximation as that for the conventional FQHE if assuming a
$p$-wave EFQH gap for the bulk CFs with the upmost Landau level
being half filled. One can replace the band mass by the
Hartree-Fock effective mass. For a pure $p$-wave superfluid, the
pairing interaction gives a finite velocity of the edge Majorana
fermionic excitation \cite{FFN}. However, in this Letter, we will
show that for the EFQH state, this conventional $p$-wave paring
interaction does not contribute to the velocity of the edge
Majorana fermionic modes. We assume the bulk states have a
$p$-wave gap. By integrating out the bulk states, the effective
theory of the edge states includes neutral chiral Majorana
fermionic modes and a charged bosonic mode which is described by
the Calogero-Sutherland model \cite{cs,suth}. Although a
conventional $p$-wave pairing interaction between edge CFs does
not contribute to the dynamics of the edge fermionic modes, there
is a bilinear coupling between the neutral and charged modes. The
velocity of the neutral modes is determined by this important
coupling. It vanishes in the ground state as naively expected but
linearly increases as the wave vector of the charged mode.

\vspace{0.1cm}

\section{General description}

A two-dimensional interacting spinless electron gas in a high
magnetic field is governed by the following Hamiltonian
\begin{eqnarray}
H&=&\sum_{\alpha=1}^N\frac{1}{2m_b}[\vec{p}_\alpha -\frac{e}{c}
\vec{A}(\vec{r} _\alpha)]^2\\
&+&\sum_{\alpha<\beta}V(\vec{r}_\alpha-\vec{r}_\beta) +\sum_\alpha
U(\vec{r}_\alpha)+U_b,\nonumber
\end{eqnarray}
where $V(\vec{r})$ is the interaction between electrons. $m_b$ is
the band mass of the electron; $U(\vec{r})$ is the external
potential trapping the electron gas in a disc and $U_b$ is the
interacting potential of the neutralizing positive background
charge \cite{cw}. The CF theory is a very useful tool in the FQHE
physics \cite{jain,HRL}. We begin with the CF transformation which
reads $
\Phi(z_1,...,z_N) = \prod_{\alpha<\beta}\biggl[\frac{z_\alpha-z_\beta} {%
|z_\alpha-z_\beta|}\biggr]^{\tilde\phi} \Psi(z_1,...,z_N),
$
where $\Phi$ is the electron wave function and $\tilde\phi$ is an
even integer. We assume the bulk states have a gap which is caused
by a $p$-wave pairing of CFs for a filling factor
$\nu=1/\tilde\phi$ and all gapless excitations are in the edge.
 We now would like to study the
effective theory of the CF edge excitations in a disc. The
partition function of the system is given by
\begin{eqnarray}
Z&=&\sum_{N^e} C^{N^e}_{N}\int_{\partial} d^2z_1....d^2z_{N^e}
\int_{B}d^2z_{N^e+1}...d^2z_N \\
&\times&\biggl(\sum_\delta |\Psi_\delta|^2 e^{-\beta
(E_\delta+E_g)}+\sum_\gamma|\Psi_\gamma |^2e^{-\beta
(E_\gamma+E_g)}\biggr),\nonumber
\end{eqnarray}
where $N_e$ is the CF number in the edge and $N$ is the total electron
number. We have divided the sample into the edge $\partial$ and the bulk $B$. $%
E_g$ is the ground state energy and $E_\delta$ are the low-lying
gapless excitation energies with $\delta$ being the excitation modes index. $%
E_\gamma$ are the gapped excitation energies. One can integrate
over the gapped state and the partition function of the system may
be written as
\begin{eqnarray}
Z&\simeq&\sum_{\delta, N^e}C_N^{N^e} \int_{\rm edge}
d^2z_1...d^2z_{N^e}
|\Psi_{e,\delta}|^2 e^{-(\beta E_\delta(N^e)+E_g)} \nonumber\\
&=&\sum_{N^e}C_N^{N^e}{\rm Tr_{\rm (edge)}}e^{-\beta H_e}.
 \label{apf}
\end{eqnarray}
In terms of the
partition function (\ref{apf}), there is the most probable edge CF number $%
\bar{N}^e$ which is determined by $\delta Z/\delta N^e=0$.
$\bar{N}^e=\int dx \rho_e(x)$ with the edge density
$\rho_e(x)=h(x)\bar n$ . Here $h(x)$ is the edge deformation and
$\bar n$ is the average density of the bulk electrons. We do not
distinguish $\bar{N_e}$ and $N_e$ hereafter if there is no
ambiguity. We do not study the bulk physics in the present work
and assume the CF interaction has been renormalized to a weak one.
The electron band mass has been renormalized to $m^*$, the
effective mass of the CF, which is finite with the order of the
Coulomb interaction as we has explained. Hereafter, we use the
unit $ \hbar=e/c=2m^*=1$.
 For the disc
sample with a radius $R$, the edge CFs are restricted in a
circular strip near the boundary with its width $\delta
R(\vec r)\ll R$ . The edge Hamiltonian of CFs reads
\begin{eqnarray}
H_e&=&\sum_{i=1}^{N^e}[\vec{p}_i-\vec{A}(\vec{r}_i) )
+\vec{a}_e(\vec{r}_i)+\vec{a}_b(\vec{r}_i)]^2 \nonumber\\
&+&\sum_{i<j}V_{eff}(\vec{r}_i-\vec{r}_j)+ \sum_i
U_{eff}(\vec{r}_i),
\end{eqnarray}
where $V_{eff}$ is the effective interaction between edge CFs and
$U_{eff}$ is the effective trapping potential including the
interaction between the edge and bulk particles. The detailed
expression of $U_{eff}$ may be quantitatively important in
numerical calculation \cite{cw}, but here for simplicity, we
suppose the trapping potential is an infinite wall for $r\geq R$
in order to analytically study a sharp edge state. Although this
may not correctly reflect the quantitative behavior, the
qualitative property which we are studying would not be changed.
In real samples, the trapping potential is very dependent on the
sample cleaving. If $U_{eff}$ is not so sharp, the edge
reconstruction is inevitable. In this case, more branches of the
edge excitations may appear \cite{wan}. The statistics gauge field
$\vec{a_e}$ is given by
\begin{eqnarray}
&&\vec{a}_e(\vec{r}_i)+\vec{a}_b(\vec{r}_i)=\frac{\tilde\phi}{2\pi}\sum_{j\not{=}i} \frac{\hat{z}%
\times (\vec{r}_{i}-\vec{r}_{j})}
{|\vec{r}_{i}-\vec{r}_{j}|^2}\nonumber\\
&&+\frac{\tilde\phi}{2\pi}\sum_{\alpha\in {\rm bulk}}\frac{\hat{z}
\times (\vec{r}_{i}-\vec{r}_{\alpha})}
{|\vec{r}_{i}-\vec{r}_{\alpha}|^2}.
\end{eqnarray}
Taking the polar coordinate $
z_i=r_ie^{i\varphi_i}$, the vector potential $%
A_\varphi(\vec{r}_i)=\frac{B}{2}r_i$ and $A_r(\vec{ r_i})=0$
and substituting the polar variables and the vector
potential to $H_e$, one has
\begin{eqnarray}
H_e&=&\sum_i\biggl[-\frac{\partial^2}{\partial r_i^2}+ (-\frac{i}{r_i}\frac{%
\partial} {\partial \varphi_i}+
\frac{\tilde\phi(N_e-1)}{2r_i})^2\nonumber \\&+&\frac{\tilde\phi^2}{4R^2}\sum_i(\sum_{j\not{=}i} \cot\frac{\varphi_{ij}}{2%
})^2 -i\frac{\tilde\phi}{R}\sum_{i<j}\cot\frac{\varphi_{ij}}{2}
\cdot
[\frac{%
\partial}{ \partial r_i}
-\frac{\partial}{\partial r_j}]\nonumber \\
&-&\frac{1}{R} \frac{
\partial}{\partial r_i}\biggr]+V_{eff}+U_{eff}+O(\delta R/R), \label{HE}
\end{eqnarray}
where we use the mean field approximation $\bar a_{b,r}=0$ and
$\bar a_{b,r}=-\frac{B}2r_i$.

\vspace{0.1cm}

\section{Solutions}

Schr\"odinger equation $H_e\Psi=E\Psi$ holds for a
 FQH edge state with a general
filling factor.
 For a given filling factor, the solution of Schr\"odinger equation
 is dependent on the bulk states as we have seen
 in solving the equations for the odd denominator FQHE \cite{yu}.
 For an even denominator quantum Hall state, e.g., the $\nu=5/2$ state \cite{exp,xia}, numerical calculations evidenced that Moore-Read Pfaffian state may be
 the ground state\cite{Morf}.  Motivated by the bulk
 ground state, we can write the general
form of the wave function as
\begin{eqnarray}
 \Psi(z_1,...,z_{N_e})= \exp \biggl\{-i
 \sum_{i<j}\frac{r_i-r_j}{4R}\cot
\frac{\varphi _{ij}}2 \biggr\}\nonumber\\\times {\cal
A}(z_1,...,z_{N_e})f(r_1,...,r_{N_e})\phi_{cs}(\varphi_1,...,\varphi_{N_e})
 .\end{eqnarray}
 We anticipate that $f$
determines the chirality of the edge charged mode which is
described by $\phi_{cs}$. Then, ${\cal A}$ corresponds to the
neutral fermionic modes. In fact, Milovanovic and Read have
written down the wave functions of the edge excitations
\cite{MR,FFN}. We will show that our solutions are consistent with
their wave functions while we confirm that the neutral fermion
modes indeed have a linear dispersion and are chiral. The new
observation is that the velocity of this fermionic modes is
proportional to the wave vector of the charged bosonic edge mode.

In terms of the general wave function, the problem ready to solve
yields
\begin{eqnarray}
&&\sum_i\biggl\{-\frac{\partial^2 }{\partial
r_i^2}-\frac{1}R\frac{\partial }{\partial r_i}
-\frac{2}{R^2}\frac{\partial \ln \chi_{cs}}{\partial
\varphi_i}\frac{\partial {\ln\cal A}}{\partial
\varphi_i}\nonumber\\&&-2\frac{\partial{\ln \cal A}}{\partial
r_i}\frac{\partial }{\partial
r_i}+U_{eff}\biggr\}f=(E-E_\varphi)f\label{10}
\end{eqnarray}
with
\begin{eqnarray}
&&-\sum_i\frac{1}{R^2}\frac{\partial^2 }{\partial
\varphi_i^2}\chi_{cs}+\frac{1}{4R^2}\sum_{i<j}
\frac{\tilde\phi(\tilde\phi-1)}{\sin^2(\varphi_{ij}/2)}\chi_{cs}
+V_{eff}\chi_{cs}\nonumber\\&&=E_\varphi\chi_{cs},
\label{CS}\\&&(\nabla^2{\cal
A})\phi_{cs}=0,~~\sum_i-i\frac{\partial {\cal
A}}{\partial\varphi_i}=\sum_il_i{\cal A},\label{mf}
\end{eqnarray}
where
$\chi_{cs}=e^{(i/2)\tilde\phi(N_e-1)\sum_i\varphi_i}\phi_{cs}$ and
$l_i$ is a set of integers or half-integers to be determined. Eq.
(\ref{CS}) describes the charged edge excitations. If one neglects
$V_{eff}$ in Eq. (\ref{CS}), it is the famous Calogero-Sutherland
model with $E_\varphi=\sum_i \frac{n_i^2}{R^2}$, whose solutions
are $ \chi_{cs}=\Phi({\bf
n})\prod_{i<j}[\sin\frac{\varphi_{ij}}2]^{\tilde\phi} $ where
$\Phi({\bf n})$ is the Jack polynomial \cite{Jack} with the
highest weight state satisfying
\begin{equation}
n_i=I_i+\frac{1}{2}\sum_{j\not=i}(\tilde\phi-1){\rm
sgn}(n_i-n_j), \label{aba}
\end{equation}
where $I_i$ are a set of integers with respect to physical
momentum along the azimuthal direction. With the interaction
$V_{eff}$, in the limit of dilute gas,
$n_i=I_i+\frac{1}{2}\sum_{j\not=i}\theta (n_j-n_i)$ with $\theta
(\Delta n)=(\tilde\phi-1){\rm sgn}(\Delta
n)-2\delta_{\tilde\phi-1}(\Delta n)$ \cite{yu}. The phase shift
$2\delta_{\tilde\phi-1}(\Delta n)$ comes from the interaction
$V_{eff}$ which is continuous as a function of $\Delta k=\Delta
n/R$ and vanishes at $\Delta k=0$ if $V_{eff}$ is short range.
Because of the step function ${\rm sgn}(\Delta n)$, the short
range interaction may not affect the low-lying behavior of the
edge modes. However, the velocity of the charged mode is lifted by
the interaction \cite{yu,yu1}.( For details, see Refs.
\cite{yu1}.) The Coulomb interaction may cause an additional
charged branch of the edge excitations with a dispersion $\Delta
k\ln \Delta k$ \cite{yu}, which is precisely the one-dimensional
plasmon excitation caused by the Coulomb interaction \cite{wen1}.
Now, we phenomenologically assume $V_{eff}$ includes a
conventional $p$-wave pairing interaction
\begin{equation}
H_{pair}=i\sum_i\Delta\cdot\left(
\begin{array}{cc}
0 & \bar\partial_i \\
\partial_i& 0
\end{array}
\right),
\end{equation}
where $2\times 2$ matrix acts on a two-components of the complex
spinless CF state \cite{FFN}.
 For a pure $p_x+ip_y$ superconductor without the bosonic mode, this $p$-wave paring interaction gives a gapless chiral fermion excitation
with its velocity $v_n=\Delta$ at the edge \cite{FFN}. However,
for the EFQH state, the pairing interaction may only change the
velocity of the charged mode but not contribute to that of the
neutral modes since $\sum_i \partial {\cal A}/\partial z_i=\sum_i
\partial {\cal A}/\partial \bar z_i=0$ according to the total antisymmetry of
${\cal A}$ with the exchange of the CFs.

Eqs.(\ref{mf}) imply that if we can find the eigen states of the
second equation, which satisfy the first one, these states indeed
describe the neutral Majorana fermion modes with a linear
dispersion. The Moore-Read Pfaffian state Pf$(\frac1{z_i-z_j})$
satisfies these equations with
 $\sum_il_i=-2$. This is the
ground state for even edge CF number. The ground state with odd
edge CF is given by $ \psi_l={\cal
A}\biggl(z_1^0\frac{1}{z_2-z_3}...\frac{1}{z_{N_e-1}-z_{N_e}}\biggr).
$ The multi Majorana fermion excited states are given by
\cite{MR,FFN}
\begin{eqnarray}
{\cal A}&=&{\cal
A}\biggl(z_1^{l_1}...z_s^{l_s}\frac{1}{z_{s+1}-z_{s+2}}...\biggr)
\nonumber\\
&=&{\cal A}\biggl(\sum
\epsilon_{i_1i_2...i_s}z_{i_1}^{l_1}z_{i_2}^{l_2}...
z_{i_s}^{l_s}\psi_{i_1i_2...i_s}^{i_{s+1}...i_N}\biggr),\label{15}
\end{eqnarray}
where $\psi_{i_1i_2...i_s}^{i_{s+1}...i_N}$ is a product of
$\frac{1}{z_i-z_j}$ where $z_i$ and $z_j$ do not include
$z_{i_1},z_{i_2},... ,z_{i_s}$ and of course is symmetric for any
permutation of $i_1,i_2,...,i_s$.

\vspace{0.2cm}

\section{The dynamic origin of the neutral fermion velocity and
chirality of the charged mode}

We now study the dynamics of the neutral fermion and chirality of
the edge excitations. As we have seen, the requirement of
$(\nabla^2{\cal A})\phi_{cs}=0$ gives rise to the chirality of the
neutral Majorana fermion edge modes. However, what is the origin
of the motion of the neutral modes for $V_{eff}$ does not
contribute to $v_n$? On the other hand, the charged mode at the
half-filling do not feel an effective magnetic field. The
Calogero-Sutherland model has both left- and right-moving gapless
modes from the Fermi points. Why is the edge charged mode still
chiral?  To answer these questions, we consider the radial
equation (\ref{10}). The pseudo-momentum of the
Calogero-Sutherland model is defined by $k_i=n_i/R$. The
pseudo-Fermi points are in $\pm k_F=\pm\tilde\phi\bar N_e/2R$ and
low energy excitations are around $k\sim \pm(k_F+q)$ for $q\ll
k_F$. We define the neutral fermion momentum $p_i=l_i/R$ and
consider multi Majorana fermion modes given by Eq.(\ref{15}).
Notice that $-i\frac{\partial \chi_{cs}}{\partial
\varphi_i}=\sum_P k_{P_i} \chi_P$ and $-i\frac{\partial {\cal
A}}{\partial \varphi_i}={\cal A}\biggl(\sum_{i_1i_2...i_s}
\epsilon_{i_1i_2...i_s}(\delta_{i_1 i}l_1+...+\delta_{i_s
i}l_s)z_{i_1}^{l_1}z_{i_2}^{l_2}...
z_{i_s}^{l_s}\psi_{i_1i_2...i_s}^{i_{s+1}...i_N}\biggr)$, where
$P$ is a permutation of $\{1,...N_e\}$. The fourth term in Eq.
(\ref{10}) reads
\begin{eqnarray}
-\frac{2}{R^2}\sum_i\frac{\partial\ln \chi_{cs}}{\partial \phi_i}
\frac{\partial\ln {\cal A}}{\partial \phi_i}=2\sum_{a=1}^sp_a
\sum_i k_i C_{k_ia}
\end{eqnarray}
where
\begin{eqnarray}
C_{k_ia}&=&\frac{1}{\chi_{cs}{\cal A}}{\cal A}\biggl(\sum
_{i_1,..,i_a,...,i_s}
\epsilon_{i_1...i_a...i_s}z^{l_1}_{i_1}...z_{i_a}^{l_a}...z^{l_s}_{i_s}\nonumber\\
&\times&\psi_{i_1i_2...i_s}^{i_{s+1}...i_N}
\sum_{\{P|P_{i_a}=i\}}\chi_P\biggr) \nonumber\\\end{eqnarray} The
symmetry of the reflection gives that $C_{k a}=C_{-ka}$. Thus, we
have $\sum_ik^{(0)}_iC_{k_i^{(0)}a}=0$ for the ground state
$\{k^{(0)}_i\}$. The low-lying excitations are given by
$\{k^{(\pm)}_i\}=\{\pm(k_F+q),k^0_i\ne  \pm k_F\}$ for $q>0$. Thus
the radial equation reads
\begin{eqnarray}
&&\sum_i\biggl\{-\frac{\partial^2 }{\partial r_i^2}+2k_Fq \pm
\sum_a 2qC_{\pm k_F a}p_a \nonumber\\&&-2\biggl(\frac{\partial
\ln{\cal A}}{\partial r_i}\biggr)\frac{\partial }{\partial
r_i}+\frac{1}R\frac{\partial }{\partial
r_i}+U_{eff}\biggr\}f=E^\pm f.\label{11}
\end{eqnarray}
The third term in Eq. (\ref{11}) is a bilinear coupling between
the wave vectors of the neutral and charged modes. Since $C_{-k_F
a}=C_{k_Fa}$ which is assumed to be positive, this coupling means
that the charged mode around $k_F$ is accompanied by a set of
neutral modes while the excitations around $-k_F$ is not physical
because the accompanied neutral fermion modes lowered the energy
of the system so that it becomes not low bounded. The physical
excitations are confined around $k_F$. This proves the chirality
of the charged edge bosonic mode with its velocity $v_s=v_F=2k_F$.
The neutral modes have velocities $v_{n,a}=2C_{k_Fa}q$. This is an
important observation made in this work : $v_{n,a}\ll v_s$ is
linearly dependent on the wave vector of the charged mode. This
dynamic origin of the velocity of the edge Majorana fermion was
never expected in the existed literature.

 We emphasize that the origin of the chirality of the charged
edge mode at the half-filling factor is very different from that
for the FQHE in the odd denominator filling factor. For the
conventional FQHE, the edge CFs see an effective magnetic field
and the CFs' cyclotron motion in this effective magnetic field is
the origin of the chirality. The CFs in a half-filling factor do
not see such an effective magnetic field. The chirality of the
charged bosonic edge mode stems from the coupling of this mode
with the neutral Majorana chiral fermion modes. Of course, from
the point of view of electron motion, the chirality of the edge
excitations is still caused by the magnetic field. This is
reflected in the wave functions: If we reverse the magnetic field,
$z$ should be replaced by $\bar z$ in all wave functions, which
leads to a reverse of the chirality of the edge excitations.

\vspace{0.1cm}

\section{ Conclusions}~~We have studied the dynamics of the edge
Majorana fermions for the EFQH state with $\nu=5/2$. We found that
there is a bilinear coupling between neutral and charged modes in
the effective edge theory, which determines the dynamics of the
neutral modes and the chirality of the charged modes. The velocity
of the neutral fermion  is proportional to the wave vector of the
charged mode. The chirality of the charged edge mode origins from
its coupling with the neutral modes.

\section{Acknowledgement}

The author is grateful to H. Y. Guo, X. Wan, Z. H. Wang, K. Yang
and M. Yu for useful discussions. This work was supported in part
by Chinese National Natural Science Foundation.

\end{document}